# Spinning Disk - Remote Focusing Microscopy

MICHELE GINTOLI[1], SHARIKA MOHANAN[1], PATRICK SALTER[3], ELIZABETH WILLIAMS[2], JAMES D. BEARD[2], GASPAR JEKELY[2], ALEXANDER D. CORBETT[1*].

1. Department of Physics and Astronomy, University of Exeter, EX4 4QL, UK
2. Living Systems Institute, University of Exeter, EX4 4QD, UK.
3. Department of Engineering Science, University of Oxford, Parks Road, Oxford, OX1 3PJ, UK

*a.corbett@exeter.ac.uk

**Abstract:** Fast confocal imaging was achieved by combining remote focusing with differential spinning disk optical sectioning to rapidly acquire images of live samples at cellular resolution. Axial and lateral full width half maxima less than 5 µm and 490 nm respectively are demonstrated over 130 µm axial range with a 256 × 128 µm field of view. A water-index calibration slide was used to achieve an alignment that minimises image volume distortion. Application to live biological samples was demonstrated by acquiring image volumes over a 24 µm axial range at 1 volume/s, allowing for the detection of calcium-based neuronal activity in *Platynereis dumerilii* larvae.

## 1. Introduction

A key challenge of fluorescence microscopy is to capture the complex three-dimensional behaviour of living organisms at high spatial and temporal resolution. Confocal microscopy provides a means of obtaining thin optical sections to study slow morphological changes in three dimensions. However, the point scanning geometry of confocal microscopes severely limits the rate at which images can be acquired. Fast point scanning technologies such as acousto-optic deflectors (AODs) can be implemented to achieve frame rates of 39 Hz or more at 512 x 512 resolution with a Z range of several hundred microns [1]. The compromise for AOD based technologies is that the short dwell time these scan rates require either limits the signal to noise ratio or requires a high light dose to the sample. Other approaches have combined AODs with electro-tuneable lenses (ETL) to change objective focus [2]. Whilst this approach was capable of imaging two planes in quick succession (3 Hz/plane) the ETL limited the effective NA of the imaging objective and was also not conjugated to the pupil plane of the objective, leading to depth-dependent imaging artefacts as described in [3]. Alternatively, piezo-actuated imaging objectives allow the focal plane to be moved through the sample at a few 10s of Hz [4] but at the cost of disturbing the sample.

Remote focusing is a method that allows rapid refocusing without moving the sample or objective. The method entails producing a faithful three-dimensional image of the sample volume at the focus of a second matched objective [5]. By accurately mapping spatial frequencies from the pupil plane of the imaging objective onto the same spatial frequencies in the pupil plane of the refocusing objective, it is possible to cancel out aberrations (predominantly spherical) generated when imaging outside of the focal plane. A third objective can then be used to relay images from this stigmatic image back to a scientific camera without the risk of sample agitation or the loss of numerical aperture.

More recently, light sheet microscopy has provided the means of acquiring optical sections at a rate limited only by the fluorophore and the frame rate of the area detector. Single objective light sheet in particular has been effective at producing single optical sections when combined with remote focusing microscopy. Here the imaging objective provides both the light sheet excitation (via pencil beam illumination at the edge of the pupil) and the widefield fluorescence detection. An aberration-free image of fluorescent emission from the light sheet can then be produced away from the specimen, at the focus of the refocusing lens. Locating the light sheet image in the focal plane of a final reimaging objective allows for *en face* wide field imaging of the light sheet [6–9].

A key challenge in single objective light sheet has been to maintain the numerical aperture of the system, such that the imaging objective remains the limiting aperture. The off-axis orientation of the final reimaging objective initially made this difficult until the introduction of a glass index solid immersion objective with a front surface cut to match the plane of the light sheet image, thereby allowing the full numerical aperture of the reimaging objective to be maintained [10–12].

In this paper we introduce an optical design that does not require any off-axis optics, but instead uses structured illumination via a spinning disk (SD) to achieve optical sectioning. SD microscopes are able to acquire optical sections at a much higher rate than standard confocal systems for a small compromise in axial resolution and/or optical efficiency. Whilst in principle any form of structured illumination can be used, the use of an off the shelf spinning disk system in the spinning disk - remote focusing (SD-RF) microscope minimises complexity whilst retaining sufficient configurability to cover most imaging tasks.

**2. Methods**

*2.1 SD-RF imaging principle*

The SD-RF is composed of three components (Fig. 1); the microscope stand, a custom remote focusing unit and the spinning disk. An image of the structured illumination pattern from the SD is formed in the focal plane of the reimaging lens which can be axially translated via a piezo, thereby controlling the location of the structured illumination pattern relative to the focal plane of the refocusing lens. As the refocusing lens is conjugated to the imaging lens, the piezo also determines the depth at which the structured illumination pattern is formed in the sample. Owing to the symmetry of the optical system, fluorescent emission from the axially translated focal plane is then reimaged back through the remote focusing system and focused in the plane of the spinning disk. The fluorescent light then undergoes spatial filtering before detection with a scientific camera (details below).

*2.2 The microscope stand*

The microscope stand is an inverted Olympus IX73, to which was attached a 40X 0.8 NA dipping objective (LUMPFLN40XW, Olympus). As the location of the back focal plane of the imaging objective remained fixed, the sample was moved into the focal plane of the lens. A manual Olympus translation stage allowed lateral sample movement, whilst axial movement was provided by a single axis micrometer (PT1 25 mm, Thorlabs) mounted firmly onto the Olympus stage and attached to a custom-made sample holder. Fluorescence captured from the dipping objective is focused by an internal 180 mm focal length lens (L1, Fig. 1) and forms the first image at a fixed distance from the side port.

*2.3 The RF unit*

Light from the side port of the inverted microscope was collected by a 1" diameter, 140 mm tube lens (L2 in Fig. 1, G063235000, Qioptiq Ltd.), housed in an XY translation mount (CXY2, Thorlabs). This mount was attached via cage rods to a cube mount containing an optical flat (DM2, Fig. 1) with a broadband, visible spectrum coating (Chroma). Lateral displacement adjustment of L2 together with tip and tilt adjustment on DM2 allowed for complete control of beam centre and tilt angle on entry into the back focal plane of the refocusing objective in the RF unit. Two 40X 0.95 NA air objectives (UPlanSApo40X2, Olympus) were mounted vertically and arranged nose to nose. As these objectives were coverslip corrected, 170 µm thick, 12 mm diameter coverslips were glued at the front surface of the objectives. The reimaging objective was attached to a piezo translator (P-725K085 PIFOC, Physik Instrumente) to allow for refocusing. Finally, an achromatic 180 mm tube lens (L3 in Fig. 1, #36-401, Edmund Optics) was used to reimage the focal plane of the reimaging lens onto the spinning disk. To meet the criteria for stigmatic imaging [5], the magnification of the tube lenses would need to be equal to the ratio of immersion media, i.e. 1.33 for a water immersion sample and air spaced remote focusing objective. The tube lens in the Olympus microscope, L1, has a focal length of 180 mm, requiring L2 to be (180/1.33) = 135 mm. The nearest off the shelf focal length for a 1" optic (140 mm) was chosen for L2.

*2.4 The spinning disk*

The spinning disk, (Clarity, Aurox Ltd.), is based on a quartz spinning disk which is imprinted with a reflective binary grating pattern (using the principles described in [13]). The disk sits in a plane conjugate to the sample, and is illuminated by an LED array (CoolLED, p300). Excitation light passes through the grating on the disk and is re-imaged through the RF system and microscope onto the sample. Translation of the reimaging lens axially displaces the location of the structured illumination pattern in the sample. Fluorescent emission from the sample is re-imaged back through the RF system (to de-scan in Z) and onto the disk. The disk is inclined at a small angle to the beam axis to allow collection of both reflected and transmitted fluorescent emission. Light emanating from the focus of the structured illumination pattern is imaged back onto the disk and passes through the apertures in the grating. This light is deflected by dichroic mirrors and focused onto one half of the camera chip to form one half of the raw image. Defocused fluorescence from outside of the focal plane is largely reflected from the binary grating to form the second half of the raw image. Using a weighted subtraction of the 'in-focus' and 'out-of-focus' images, it is possible to obtain high contrast, optically sectioned images. To minimise non-common path errors and ensure best performance, the two halves of the image are manually aligned using an internal calibration pattern and custom software.

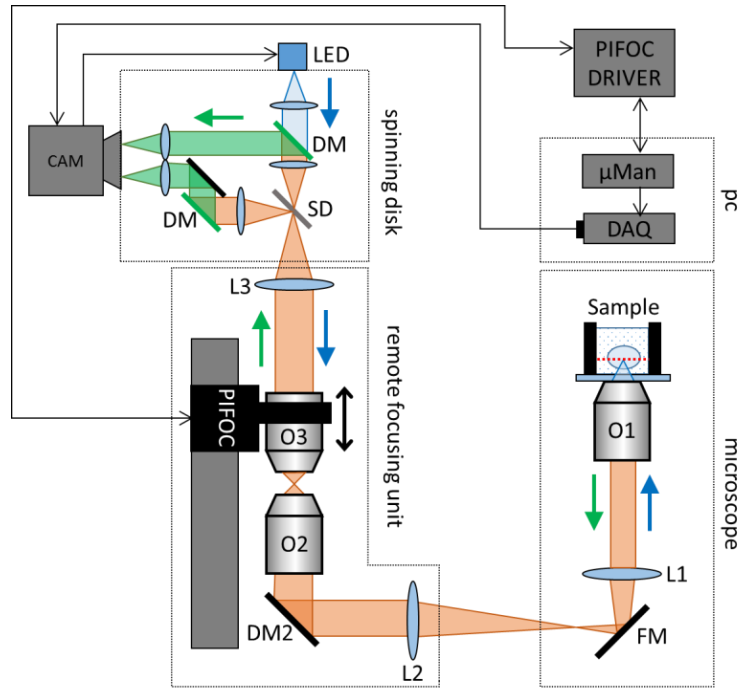

Fig. 1: Simplified optical scheme of the combined SD-RF setup: Structured excitation light from the spinning disk (SD) is imaged through the remote focusing unit and onto the sample. Sample fluorescence is collected by O1 (40X 0.8NA, water immersion) and the magnified image is demagnified by the refocusing lens, O2 (40X 0.95 NA air). A third reimaging objective O3, identical to O2, relays an image of the sample plane, back onto the spinning disk (SD). In-focus fluorescence passes through the upper path onto one half of the sCMOS camera (CAM), with the out of focus component passing through the lower path onto the second half of the camera detector. On-the-fly processing of the two images then returns a confocal image of the sample. By scanning and de-scanning in Z, the position of O3 determines the axial location of plane of interest in the sample (see main text). FM=fold mirror, DM2=dischroic mirror. Tube lenses L1 (f=180 mm) and L2 (f=140 mm).

Higher spatial frequency gratings correspond to thinner optical sections, at the cost of lower 'in-focus' signal. The spinning disk is imprinted with three different grating frequencies to cover the trade-off between high-signal, low sectioning to low-signal, high sectioning. The disk also has a 'bypass' option to allow for widefield imaging.

*2.5 Synchronisation*

Image detection was performed using a sCMOS camera with rolling shutter (Zyla 4.2, Andor Technology, Oxford Instruments). The need to capture both in-focus and out-of-focus images on the camera halved the native image resolution to 2048 × 1024 pixels. Camera frame buffer acquisition, frame exposure time, frame delay, PIFOC drive signal and LED illumination were controlled by MicroManager.

To provide a more accurate frame timing than the default camera trigger, pulses were generated from a data acquisition card (PCIe-6363, National Instruments) an external trigger, giving a constant frequency (in the range 1-20 Hz) with a jitter ≤ 1ms. To minimise specimen exposure and avoid unnecessary photobleaching, the LEDs were triggered to switch on only during frame acquisition, when the "Fire All" TTL output of the camera was in the 'on' state.

*2.6 System configurations*

The performance of the complete SD-RF system was compared to two other configurations. First, the "RF-only" configuration replaces the spinning disk with the Zyla sCMOS camera to capture remotely refocused wide field images. Second, the "side port" configuration has the camera attached to the side port of the microscope to measure the intrinsic performance of the sample and imaging lens combination. The "side-port" configuration then provides a base line performance against which the other configurations can be measured.

*2.7 Water-index fluorescent calibration sample*

To calculate the lateral and axial magnification of the microscope, a laser-written fluorescent calibration sample ("Waterslide") was used. The Waterslide was fabricated in a similar manner to [14] but used a low index polymer substrate. Briefly, a thin (250 µm) film of low index polymer (n=1.34) was cut into a 7 mm x 7 mm square and fixed to a microscope slide using an adhesive plastic washer with a central clear aperture of 5 mm diameter. The sample was mounted onto a high precision (50 nm repeatability) 3D translation stage (Aerotech ABL10100). This central region of the sample was then irradiated using an amplified Ti:Sapph laser (790 nm, 250 fs, 1 kHz rep rate, Solstice, Spectra-Physics), which was intensity modulated using a motorised half wave plate. Structures were fabricated into the polymer using a burst of five pulses for each feature.

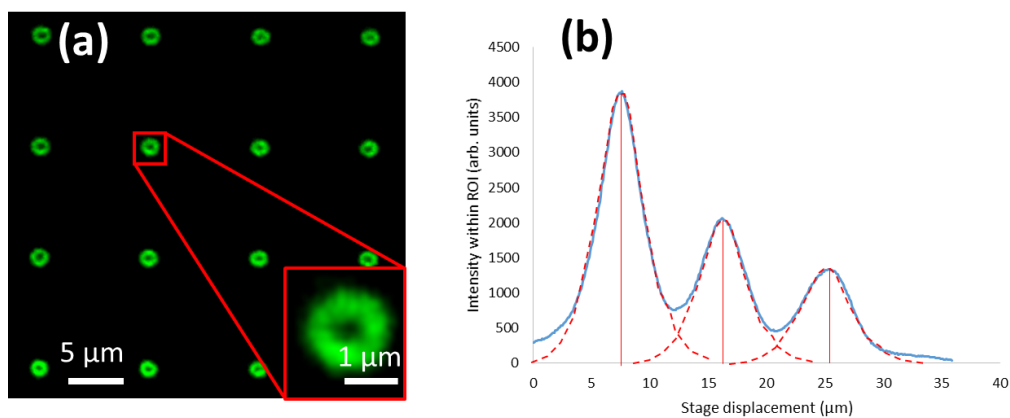

Fig. 2: (A) A fluorescent calibration pattern written into low index polymer (n=1.34) showing first layer in a 3D array of features. (B) Axial intensity profile (blue line) within an ROI encircling a single fluorescent feature in the array. The average axial separation of 8.76 µm between layers was measured by fitting a Gaussian (red dashed line) to the axial intensity profile across each layer and calculating the separation of the Gaussian peaks.

As the surface of the low index polymer is in direct contact with the water immersion medium, it avoids generating spherical aberration when imaging the fluorescent features with the 40X 0.8 NA dipping lens. The fluorescent pattern written into the Waterslide was an array of 10 × 10 × 3 (XYZ) fluorescent features. The lateral spacing between features is 10 µm and the axial separation between layers is 10 µm *(1.34/1.52) = 8.82 µm. This foreshortening of the layer separation is due to the features being written with an oil lens into a lower index medium. The lateral FHWM of the larger (and brighter) donut-shaped feature is approximately 1.5 µm (Fig. 2 (a), inset). The axial separation of the layers was confirmed by first acquiring image stacks of the entire feature array on a standard confocal (Leica TCS5) using the same 40X dipping objective. After image stacks were acquired, the axial intensity profile within an ROI containing a single fluorescent feature in the array was calculated. Three Gaussian functions were then fit to the axial intensity profile, using the fitted Gaussian peak values to calculate the axial separation of the planes (Fig. 2(b)). Repeating this procedure for each feature in the array (N=100), an average separation value of 8.76 ± 0.1 µm was calculated. Monitoring lateral shift in the feature centroid over the axial scan range demonstrated that the sample tilt was < 1°.

To calculate the lateral image magnification, an image of a single layer of features was first captured. The centroid of these features were determined using a custom image processing script (MATLAB) and the mean centroid separations calculated in pixels. Multiplying the mean separation by the 6.5 µm camera pixel size gave the physical feature separation on the camera. This could then be compared to the known 10 µm lateral separation in the sample to produce the lateral image magnification. The axial magnification was calculated as the layer separation measured by the stage relative to the known value of 8.76 µm. When imaging through the RF unit, there is an anticipated lateral and axial image magnification of $n_1/n_2 = 1.33$ [5]. This would deliver an expected apparent axial separation of 1.33 * 8.76 µm = 11.65 µm. The actual layer separation was measured from image stacks using the same method described above (Fig. 2 (b)).

*2.8 PSF measurements*

To measure the resolving power of each of the three microscope configurations, we imaged a sample composed of 100 nm fluorescent beads with excitation and emission peaks at 505 nm and 515 nm respectively (F8803, ThermoFisher) dispersed in 2% agarose. Image stacks of the beads were captured over an axial range of 400 um. For the "side port" configuration, the sample was axially translated using a single axis piezo (Q-545.140, Physik Instrumente) in 1 µm steps to allow the focal plane measurement of the lateral and axial resolution. The ImageJ plugin, PSFj (Koppen lab), was used to analyse the bead images and perform a statistical analysis of the average lateral and axial PSF widths.

### 3. Results

*3.1 Magnification measurements*

The "side port" configuration was first used to measure the lateral magnification and axial scaling. The mean lateral feature separations on the Waterslide were measured to be 61.6 pixels in both X and Y [N=81]. Given the 6.5 µm pixel size of the Andor Zyla sCMOS, this corresponds to a 400.5 µm feature separation in the detector plane and a 40.05X magnification, which is in line with the 40X expectation (Table 1).

For the SD-RF configuration, the magnification increases due to the combination of tube lenses (L1 and L2, Fig. 1). The expected lateral magnification then becomes (L1/L2)*M = (180/140)*40X = 51.42X. The results shown in Fig. 3 indicate that at the focal plane of the imaging objective (at Z=160 µm) the magnification is 51.1X. For a non-telecentric remote focusing system, a depth-dependent magnification artefact is observed ([3]). In an ideal system, where the back focal plane of the imaging objective is accurately imaged onto the remote focusing object, there should be no depth dependent magnification. In this system we observe that the change in magnification over a 400 μm depth range is less than 3%.

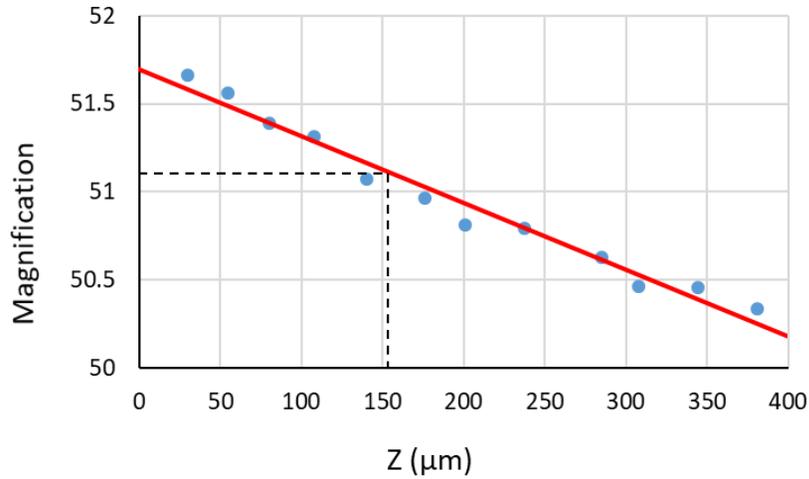

Fig. 3: Variation in lateral magnification as a function of refocusing depth for the SD-RF system. The change in magnification was < 3% over a 400 μm range. Data (blue dots) and line of best fit (red line) are shown. Dashed black lines indicate a magnification of 51.1 at the focal plane of the imaging objective.

**Table 1: Magnification summary for different microscope configurations**

| Configuration | Magnification (XY) | | Scaling (Z) | |
|---|---|---|---|---|
| | Min. | Max. | Min. | Max. |
| Side port | 40.0 X | 40.1 X | 1 X | 1 X |
| SD-RF | 50.3 X | 51.6 X | 1.26 X | 1.28 X |

*3.2 PSF measurements*

Lateral and axial PSF FHWM across the axial scan range are shown in Fig. 4 and summarised in Table 2. A significant variation can be observed over a range of 200 μm, around the Z = 50 μm position. Beyond 200 μm the resolution measurements made by PSFJ become more variable. This is due to the onset of significant astigmatism, which makes it difficult for the PSFJ software to accurately fit lateral and axial Gaussian profiles. The astigmatism is most likely due to warping of a large folding mirror inside the RF unit. The PSF measurements made at the side port of the microscope, in the focal plane of the imaging objective, are shown as the grey solid line in Fig. 4. This shows that (i) the resolution of the RF-only configuration is equivalent to that of the SD-RF system when using the "high signal" (low spatial frequency) pattern on the disk and that (ii) the axial SD-RF resolution actually exceeds that measured at the side port. This latter point is considered in more detail in the Discussion.

**Table 2: PSF dimensions measured in each microscope configuration**

|  | PSF FWHM XY (min) / nm | PSF FWHM Z (min) / µm |
| --- | --- | --- |
| Theoretical value | 330 | 1.88 |
| Side port | 424 ± 10 | 4.6 ± 0.2 |
| RF-only | 424 ± 11 | 3.3 ± 0.3 |
| SD-RF | 419 ± 22 | 3.2 ± 0.4 |

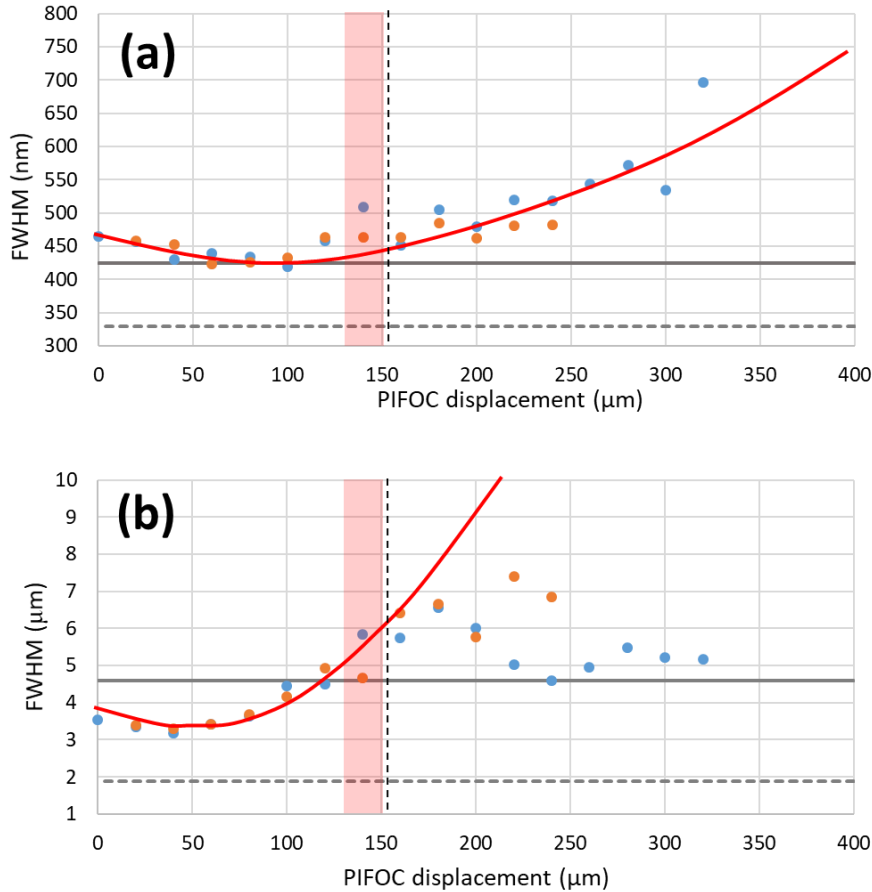

Fig. 4: Measured lateral (a) and axial (b) resolutions (FWHM) for RF-only (orange dots) and SD-RF (blue dots) imaging configurations, together with trend line (solid red line). Also shown are the imaging objective focal plane (vertical dashed black lines), theoretical resolution limits (dashed grey lines), resolution values measured at the side-port (solid grey lines) and the axial range over which live image data was taken (pink bars).

*3.3 Imaging in Platynereis dumerilii*

*Preparation of larvae for live calcium imaging:* GCaMP6-expressing *Platynereis* larvae were generated through microinjection of GCaMP6s RNA (1000 µg/µL) at one-cell stage as previously described [15]. Larvae were reared at 18°C on a 16hr light 8 hr dark cycle until the imaging experiments. Experiments were conducted between 36 and 52 hr-post-fertilization. Live larvae were mounted for imaging on glass slides and held with a coverslip spaced with adhesive tape to avoid placing excessive pressure on the larvae.

*Preparation of fixed larvae for multichannel imaging:* Immunostaining of larvae was carried out as described in [16], but with the following modifications. 2- and 3-day-old larvae were fixed with 4% paraformaldehyde in PTW (PBS + 0.1% Tween-20) for 1h at room temperature before storing in 100% methanol at -20°C. 2-day-old fixed larvae were stained with phosphorylated MAP kinase primary antibody raised in rabbit (pERK, Cell Signaling Antibody #9101, 1:500) and Alexa Fluor 647 goat anti-rabbit secondary antibody (ThermoFisher Scientific #A-21244, 1:250). During secondary antibody incubation, larvae were also stained with DAPI nuclear dye (1µg/ml). 3-day-old fixed larvae were stained with a commercial primary antibody against acetylated tubulin, raised in mouse (Sigma #T6793, 1:250) and a custom antibody against the neuropeptide THDamide (ARGGPIAAPLWFLKTHDa), raised in rabbit (1µg/ml). Secondary antibodies used for staining 3-day-old larvae were Alexa Fluor 647 goat anti-rabbit (ThermoFisher Scientific #A-21244, 1:250) and Alexa Fluor 555 goat anti-mouse (ThermoFisher Scientific #A-21226, 1:250).

For mounting following immunostaining, larvae were transferred gradually from PTW to 97% 2,2'-thiodiethanol (TDE) (166782, Sigma-Aldrich, St. Louis, USA) in steps of 25% TDE/PTW dilutions.

*3.4 Multiplane imaging*

A key advantage of the SD-RF system is the ability to rapidly image across multiple planes within a sample. To demonstrate this we set up a five frame image stack around the anterior neural plexus of a 2 day old *Platynereis* larvae. At this stage the larvae are largely spherical, with a diameter of 125-175 µm. The anterior neural plexus extends across most of the anterior hemisphere of the larva (Fig. 5(b)).

To capture the images, planes were set up at 6 µm separation starting from a depth ~15 µm below the surface of the larva, extending to a depth of ~40 µm. As shown in Fig. 5(b), the lowest position of the PIFOC (i.e. where the refocusing and reimaging objectives are closest) is labelled Z=400, which corresponds to the deepest location inside the sample. Increasing the value of Z moves the focal plane towards the coverslip. This scan began at Z=152 µm and ended at Z=128 µm. The five-stack volume was monitored for 10 seconds at a frame rate of 5 Hz (i.e. a sampling rate of 1 volume/s and 1 Hz per image plane).

The timing diagram is shown in Fig. 5(a). Frame acquisition begins with the frame trigger signal (vertical black line) sent to the camera. This included a trigger delay of up to 30 ms time before starting the 50 ms frame exposure (grey box). To minimise sample exposure, the excitation LED (bottom row) is only on whilst the rows of the frame are being exposed. Once the frame has been acquired, the next drive signal is sent to the PIFOC, leaving a maximum window of 150 ms for the translation and the settling (settling time ~ 25 ms) to occur. In total, the temporal separation between two subsequent frames is 200 ms.

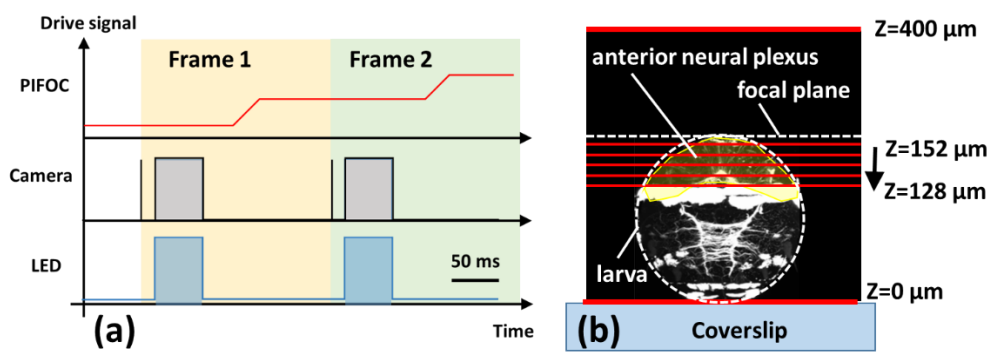

Fig. 6: (a) Timing diagram for the SD-RF during acquisition. (b). Sketch showing the locations of the acquired frames and the sample relative to the Z values used by the PIFOC. Anterior neural plexus highlighted in yellow. Image of Platynereis adapted from Fig. 1F, [17].

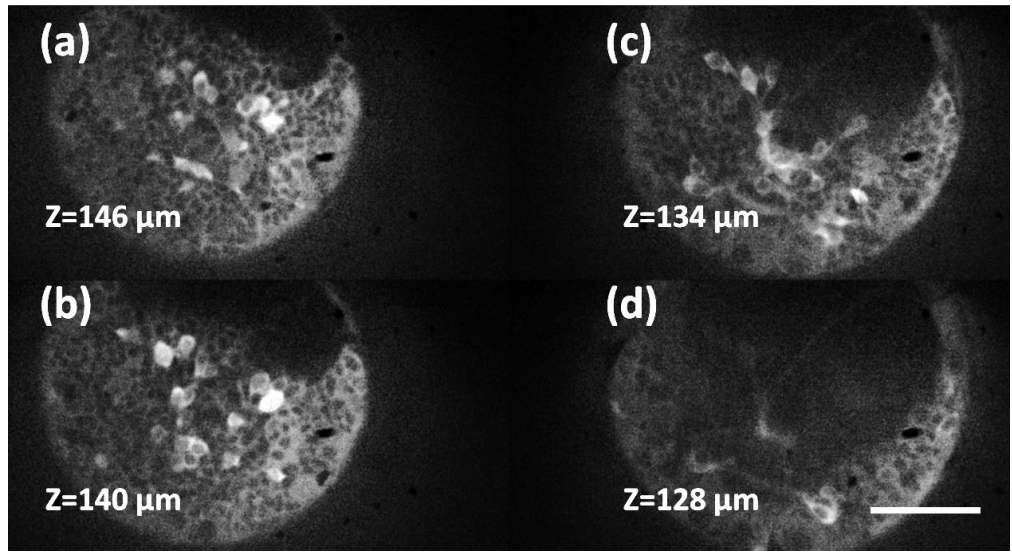

Fig. 5: Four of the five optical sections taken from the anterior nervous system region of Platynereis dumerilii (anterior view). Sections are axially separated by 6 µm over a 24 µm depth range. 50 ms exposures are captured at a rate of 5 Hz. Scale bar 50 µm.

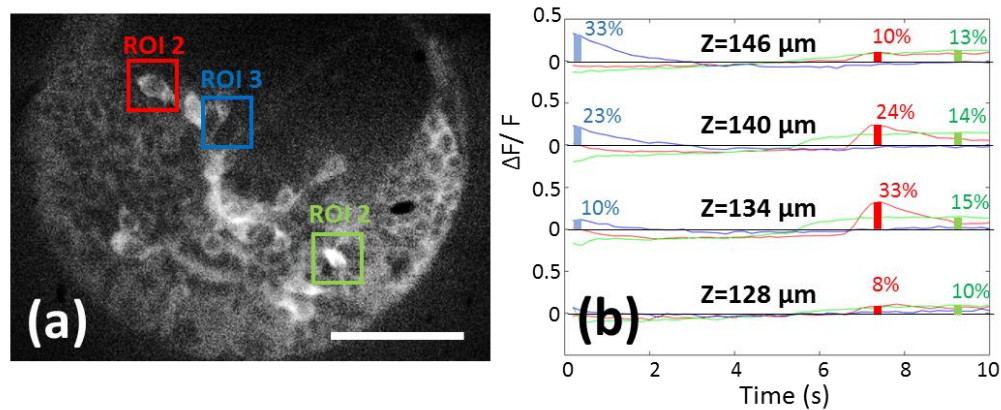

Fig. 7: Measuring changes electrical activity through $Ca^{2+}$ binding to GCaMP. Three regions of interest were defined for each of the four planes. (a) Overlays on the Z=128 µm plane show the location of the three regions of interest (ROI). Scale bar 50 µm. (b) Traces showing fluorescence changes relative to average fluorescence (ΔF/F) within the colour coded ROI across all time points. Plots are shown in vertically offset for clarity. Overlays indicate values of ΔF/F at each of the time points shown.

Four of the five image planes are shown in Fig. 6 and Visualisation 1. The SD-RF microscope was able to resolve cell bodies in each layer and detect electrical activity from normal spontaneous behaviour as the $Ca^{2+}$ flux associated with an action potential bound to the genetically expressed GCaMP (Fig. 7). To calculate the traces of electrical activity, 64 × 64 pixel regions of interest were defined (Fig. 7 (a)). The pixel values within each ROI were summed for each frame depth and at each time point. The average fluorescence (F) value in each time trace was determined and used to calculate the relative change in fluorescence (ΔF/F). Averaging over multiple traces is not required due to the low baseline noise level (ΔF/F standard deviation <0.03 for quiescent traces) thereby improving sensitivity.

*3.5 Multichannel imaging*

The spinning disk is illuminated by an LED bank (pe300, CoolLED) via a motorised filter cube (see Table 3 for a summary of the filters). It is therefore possible to remotely switch the excitation and emission bands. As this procedure can take up to 500 ms, it is more efficient to change colour channel between acquired image volumes. In live specimens, sample movement could lead to a misalignment of the different colour channels. To demonstrate the possibility of multicolour volumetric imaging in SD-RF, we imaged multi-labelled *Platynereis dumerilii* specimens that had already been fixed (Fig. 8). For better visualisation, these data sets have also been rendered as videos (see Visualisation 2 and Visualisation 3). Fig. 8 (a) shows a 2 day old larva which has been stained with DAPI and an antibody against pERK (see Section 3.3 for more details). Fig. 8 (b) shows four bands of cilia labelled with an antibody against acetylated tubulin (green) which encircle the long axis of the 3 day old larva. Also stained are the THDa neuropeptide-expressing neurons (red) with cell bodies in the anterior neural plexus and axons in the ventral nerve cord of the larva. The images acquired in Fig. 8 (a) and (b) cover volumes of 128 × 128 × 153 µm and 256 × 128 × 175 µm respectively.

**Table 3: Summary of colour filters available on the Clarity spinning disk.**

| Channel label | LED label | Excitation λ (nm) | Emission λ (nm) |
|---|---|---|---|
| DAPI | A | 370-410 | 430-475 |
| GFP | B | 446-486 | 503-548 |
| DsRed | C | 554-568 | 582-636 |
| Cy-5 | C | 626-644 | 659-701 |

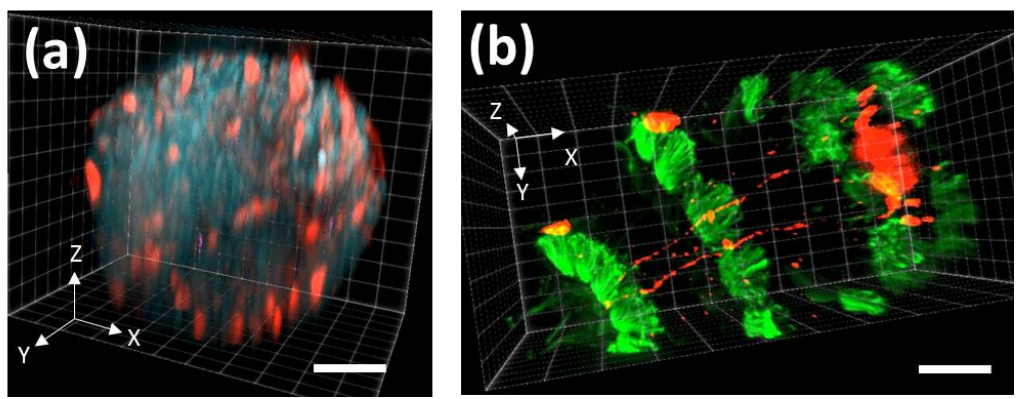

Fig. 8: Multichannel images of fixed Platynereis dumerilii larvae after (a) 2 days post-fertilisation and (b) 3 days post-fertilisation. Labels: (a) blue (DAPI) = nuclei; red (DsRed) = pERK immunostaining, (b) red = THDa immunostaining, green = cilia bands (acetylated tubulin immunostaining). Scale bars (a) 30 µm (b) 50 µm.

## 4. Discussion

*4.1 Increasing volume acquisition rate*

In this paper we allowed the PIFOC to move, stop and settle before exposing the camera frame. After all lines in the frame had been exposed, the signal was sent to the PIFOC to move to the next position. Two approaches to reduce the time taken to acquire a volume are to reduce the exposure time per frame and / or reduce the number of frames per stack. The limits on these parameters are largely determined by the biological sample. The minimum exposure time is set by the signal to noise ratio, which in turn is set by the amount of fluorophore in the sample. Whilst higher excitation intensities will increase the emissive flux, this is likely to lead to more rapid bleaching. The minimum number of frames per stack is set by the optical sectioning thickness and the particular sample features and behaviour to be observed. There is no advantage to the axial frame separation being less than half of the axial FWHM, but the minimum separation will be determined by the axial separation of the features of interest. It is anticipated that the greatest speed improvements using the SD-RF microscope, relative to a standard spinning disk microscope, is for large axial ranges where the time required for stage scanning is a large fraction of the volume acquisition duty cycle.

An alternative approach to multiplane volume acquisition is to expose the camera frame whilst the PIFOC traverses the entire axial range of the target volume. This optically sums the sectioned images on the detector, compressing the entire image volume into a 2D representation. True optical sections will be integrated in the SD-RF microscope so long as PIFOC covers a range less than the axial FWHM during the time for the spinning disk to complete one revolution. For a rotation speed of 3,000 rpm and axial FWHM of 4 µm, this corresponds to <12 mm/s axial speed or a maximum frame rate of 120 Hz over a 100 µm axial range. Extended depth of field (EDF) imaging using remote focusing was previously achieved by Botcherby et al [18] using the folded RF geometry and a Nipkow spinning disk to obtain an EDF image over a 20 µm axial range in 100 ms. Whilst this is perfectly feasible in the SD-RF system, it is not well matched to the thick tissue sample where an EDF image would obscure fluorescence changes by summing over many densely packed optical sections.

*4.2 Increasing image resolution*

The PSF FWHM measurements shown in Fig. 4 indicate that a lateral resolution of less than 500 nm was maintained over the axial range used to image *Platynereis dumerilii*, corresponding to an NA > 0.5. Compared to the 0.8 NA dipping objective used, it can be seen that there is the potential for improvement. There were three main factors that limited the achievable resolution. These were (i) spherical aberration introduced by the sample coverslip (ii) astigmatism introduced by the fold mirror in the RF unit and (iii) a magnification mismatch in the relay lenses.

The SD-RF system was characterised using a water dipping objective imaging fluorescent features in a water-index calibration sample. However, to image both live and fixed samples it was necessary to introduce a coverslip between the sample and the imaging objective. As this dipping objective was not coverslip corrected, it introduced a significant amount of spherical aberration into the imaging system. To quantify the effect of the coverslip on the achievable PSF when imaging with the dipping lens, a test sample was constructed and imaged at the side port of the inverted microscope (Fig. 9). This sample consisted of 100 nm fluorescent beads (F8803, ThermoFisher) dried onto a microscope slide. A coverslip was then sandwiched between the microscope slide and a second plastic slide containing a circular aperture. The sandwiched microscope slide occupied half of the area of the circular aperture. The sample was then inverted, water immersed and brought into the focal plane of the dipping lens. By moving the sample laterally, the dipping lens could image either the exposed beads, or those directly under the coverslip. The sample was mounted on a single axis piezo stage which was in turn fixed to the microscope stage.

The sample was illuminated using an arc lamp source. The piezo stage was stepped in 200 nm intervals and XY images acquired of the PSF. XZ projections of the PSF are shown in Fig. 9(b) with FWHM values recorded in Table 4. The data show that the introduction of a coverslip not only greatly extends the axial profile of the PSF but also shifts the location of best focus 28 µm closer to the imaging objective. This is largely due to foreshortening and the introduction of (negative) spherical aberration. The degradation of image resolution introduced by the sample coverslip can be simply avoided by using a water immersion imaging objective with coverslip correction. This would also need to have an angular aperture less than that of the 0.95 air objectives in the RF unit.

**Table 4: Lateral and axial PSF FWHM values measured at the microscope side port using a bead sample both with and without the presence of a coverslip.**

|  | Best focus Z (µm) | PSF FWHM XY (min) / nm | PSF FWHM Z (min) / µm |
| --- | --- | --- | --- |
| Theoretical value | - | 330 | 1.88 |
| No coverslip | 160 | 390 ± 3 | 1.85 ± 0.01 |
| Coverslip | 132 | 424 ± 10 | 4.62 ± 0.2 |

Of particular interest was the result reported in Table 2 which showed that the resolution reported by the SD-RF microscope is better than that measured at the side port when imaging a bead with a coverslip. This implies that the RF unit was able to compensate for the coverslip spherical aberration. This is because from the point of view of the imaging objective, the sample appears closer to the objective. Consequently, the aberrated image is pushed outside the focal pane of the reimaging lens (away from the objective). Imaging an object deeper than the focal plane introduces a positive spherical aberration by the reimaging lens which is opposite to that introduced by the coverslip. This leads to a partial cancellation of the spherical aberration before imaging.

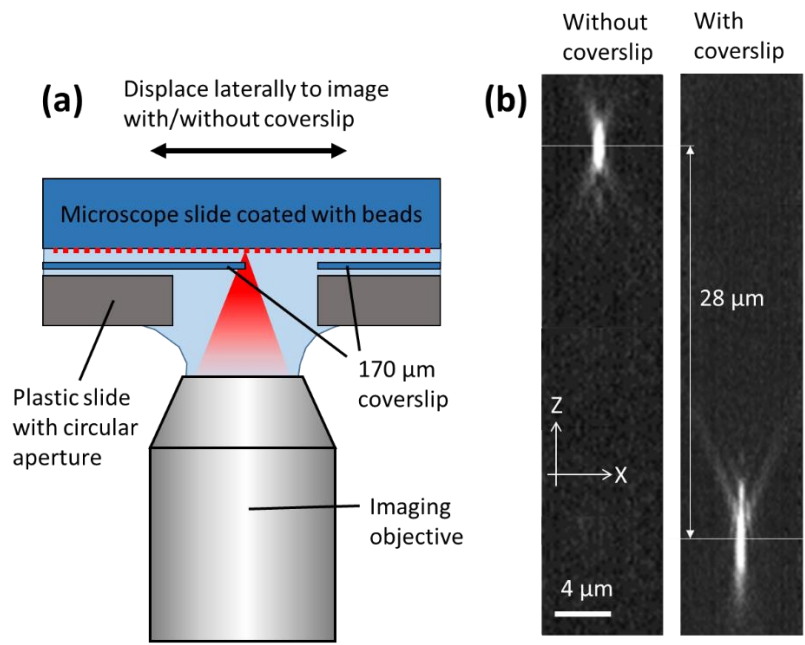

Fig. 9: (a) A test sample used to measure the PSF with and without the presence of a coverslip. A layer of dried beads on a microscope slide is half covered by a coverslip which is held in place by a second plastic slide with circular aperture. The beads can then be imaged with and without a coverslip by a small lateral displacement of the sample. (b) XZ PSF profiles obtained using the test sample in (a), showing negative spherical aberration introduced by the coverslip and the corresponding axial displacement.

In addition to spherical aberration in the SD-RF system, the image stacks taken of the 100 nm beads in agarose indicated the presence of significant astigmatism. This was most evident for Z>200 μm. As the astigmatism was seen in both the SD-RF and RF-only configurations, this was likely to be due to the 3 mm thick, broadband dichroic mirror (with coatings on both surfaces) used in the RF unit. Again, this problem can be avoided by the introduction of a prismatic fold mirror which is less susceptible to warping.

Finally, the choice of relay lenses further limited the achievable resolution. As described above, an ideal remote focusing system would have a second tube lens (L2 in Fig. 1) focal length 135 mm. The closest commercially available focal length was 140 mm. A useful resource for creating the unconventional focal length lenses used in RF systems has been created elsewhere [11]. Recent modelling has shown that even this small mismatch in magnification will both reduce the effective dynamic range of the RF unit and compromise the achievable resolution. This will be explored in more detail in a future publication.

## 5. Conclusion

In summary we have characterised the imaging performance of a combined spinning disk - remote focusing (SD-RF) microscope. Using a water index calibration sample we were able to determine that distortion of the 256 × 256 × 400 μm image volume was < 3%. Imaging 100 nm beads dispersed in agarose we showed that we could achieve axial and lateral resolutions of better than 5 μm and 500 nm respectively over an axial range of 130 μm. Multiplane imaging was demonstrated over 5 equidistant planes across the anterior nervous system of *Platynereis dumerilii*. The entire 24 μm thick image volume was sampled at a rate of 1 Hz (5 frames per second). The 256 × 128 μm images captured spontaneous electrical activity with a signal to noise ratio >10, avoiding the need to average over multiple events.

**Funding**
Engineering and Physical Sciences Research Council (EPSRC) (EP/R021252/1, EP/R004803/1). Deutsche Forschungsgemeinschaft (DFG) (JE 777/3-1).

**Acknowledgements**
We would like to thank Aurox Ltd for the loan of their Clarity spinning disk confocal that made this work possible.

**Disclosures**
The authors declare that there are no conflicts of interest related to this article.

**References**
1.   K. N. S. Nadella, H. Roš, C. Baragli, V. A. Griffiths, G. Konstantinou, T. Koimtzis, G. J. Evans, P. A. Kirkby, and R. A. Silver, "Random-access scanning microscopy for 3D imaging in awake behaving animals," Nat. Methods **13**, 1001 (2016).
2.   B. F. Grewe, F. F. Voigt, M. van 't Hoff, and F. Helmchen, "Fast two-layer two-photon imaging of neuronal cell populations using an electrically tunable lens," Biomed. Opt. Express **2**, 2035–2046 (2011).

3. A. D. Corbett, R. A. B. Burton, G. Bub, P. S. Salter, S. Tuohy, M. J. Booth, and T. Wilson, "Quantifying distortions in two-photon remote focussing microscope images using a volumetric calibration specimen," Front. Physiol. **5**, 384 (2014).
4. W. Göbel, B. M. Kampa, and F. Helmchen, "Imaging cellular network dynamics in three dimensions using fast 3D laser scanning," Nat. Methods **4**, 73–79 (2006).
5. E. J. Botcherby, R. Juškaitis, M. J. Booth, and T. Wilson, "An optical technique for remote focusing in microscopy," Opt. Commun. **281**, 880–887 (2008).
6. C. Dunsby, "Optically sectioned imaging by oblique plane microscopy," Opt. Express **16**, 20306–20316 (2008).
7. S. Kumar, D. Wilding, M. B. Sikkel, A. R. Lyon, K. T. MacLeod, and C. Dunsby, "High-speed 2D and 3D fluorescence microscopy of cardiac myocytes," Opt. Express **19**, 13839–13847 (2011).
8. M. B. Bouchard, V. Voleti, C. S. Mendes, C. Lacefield, W. B. Grueber, R. S. Mann, R. M. Bruno, and E. M. C. Hillman, "Swept confocally-aligned planar excitation (SCAPE) microscopy for high-speed volumetric imaging of behaving organisms," Nat. Photonics **9**, 113–119 (2015).
9. M. Kumar, S. Kishore, J. Nasenbeny, D. L. McLean, and Y. Kozorovitskiy, "Integrated one- and two-photon scanned oblique plane illumination (SOPi) microscopy for rapid volumetric imaging," Opt. Express **26**, 13027 (2018).
10. B. Yang, Y. Wang, S. Feng, V. Pessino, N. Stuurman, and B. Huang, "High Numerical Aperture Epi-illumination Selective Plane Illumination Microscopy," bioRxiv 273359 (2018).
11. A. Millet-Sikking and A. G. York, "High NA single objective light sheet," (2019).
12. R. Fiolka, "Resolution upgrades for light-sheet microscopy," Nat. Methods **16**, 813–814 (2019).
13. R. Juškaitis, T. Wilson, M. a. A. Neil, and M. Kozubek, "Efficient real-time confocal microscopy with white light sources," Nature **383**, 804–806 (1996).
14. A. D. Corbett, M. Shaw, A. Yacoot, A. Jefferson, L. Schermelleh, T. Wilson, M. Booth, and P. S. Salter, "Microscope calibration using laser written fluorescence," Opt. Express **26**, 21887 (2018).
15. N. Randel, A. Asadulina, L. A. Bezares-Calderón, C. Verasztó, E. A. Williams, M. Conzelmann, R. Shahidi, and G. Jékely, "Neuronal connectome of a sensory-motor circuit for visual navigation," eLife **3**, e02730 (2014).
16. M. Conzelmann and G. Jékely, "Antibodies against conserved amidated neuropeptide epitopes enrich the comparative neurobiology toolbox," EvoDevo **3**, 23 (2012).
17. A. Asadulina, A. Panzera, C. Verasztó, C. Liebig, and G. Jékely, "Whole-body gene expression pattern registration in Platynereis larvae," EvoDevo **3**, 27 (2012).
18. E. J. Botcherby, M. J. Booth, R. Juskaitis, and T. Wilson, "Real-time extended depth of field microscopy," Opt. Express **16**, 21843–21848 (2008).